\documentclass[usenatbib]{mnras}
\usepackage{natbib}
\usepackage{apjfonts}
\usepackage{amsmath}
\usepackage[table]{xcolor}

\usepackage{epsfig}
\usepackage{graphicx}
\usepackage{amssymb}
\usepackage{aas_macros}
\usepackage{color}

\newcommand{\kms}{\,{\rm km \, s^{-1}}}

\newcommand{\kpc}{\,{\rm kpc}}

\newcommand{\Gyr}{\,{\rm Gyr}}

\newcommand{\oversim}[2]{\protect{\mbox{\lower0.5ex\vbox{%
   \baselineskip=0pt\lineskip=0.2ex
   \ialign{$\mathsurround=0pt #1\hfil##\hfil$\crcr#2\crcr\sim\crcr}}}}} 
\catcode`\"=\active\let"=\"  

\def\3{{\ss} }

\def\c12{{1\over 2}}

\def\plusplus{\raise 0.3ex\hbox{${\scriptstyle ++}$}{}}

\def\and{{{\rm M}31}}

\defcitealias{pp21}{PP21}
\defcitealias{vasiliev20}{VBE21}

\usepackage{array}
\newcolumntype{L}[1]{>{\raggedright\let\newline\\\arraybackslash\hspace{0pt}}m{#1}}
\newcolumntype{C}[1]{>{\centering\let\newline\\\arraybackslash\hspace{0pt}}m{#1}}
\newcolumntype{R}[1]{>{\raggedleft\let\newline\\\arraybackslash\hspace{0pt}}m{#1}}

\begin{document}
\title[The Sgr stream in angular momentum]{Identification of Sagittarius stream members in Angular Momentum space with Gaussian mixture techniques}
\author[Pe\~{n}arrubia \& Petersen]{Jorge Pe\~{n}arrubia$^{1,2}$\thanks{jorpega@roe.ac.uk}, Michael S. Petersen$^{1}$\\
  $^1$Institute for Astronomy, University of Edinburgh, Royal Observatory, Blackford Hill, Edinburgh EH9 3HJ, UK\\
  $^2$Centre for Statistics, University of Edinburgh, School of Mathematics, Edinburgh EH9 3FD, UK
}

\maketitle  

\begin{abstract}
 This paper uses Gaussian mixture techniques to dissect the Milky Way (MW) stellar halo in angular momentum space. Application to a catalogue of 5389 stars near the plane of the Sagittarius (Sgr) stream with full 6D phase-space coordinates supplied by Gaia EDR3 and SEGUE returns four independent dynamical components.
The broadest and most populated corresponds to the `smooth' MW halo. The narrowest and faintest contains 40 stars of the Orphan stream. We find a component with little or no angular momentum likely associated with the GSE substructure.
We also identify 925 stars and 7 Globular Clusters with probabilities $>90\%$ to be members of the Sgr stream. Comparison against $N$-body models shows that some of these members trace the continuation of the leading/trailing tails in the Southern/Northern hemispheres. The new detections span $\sim 800^\circ$ on the sky, thus wrapping the Galaxy {\it twice}.  
\end{abstract}


\section{Introduction}\label{sec:intro}
Since its serendipitous discovery by \citet{ibata94}, the Sagittarius (Sgr) dwarf galaxy has become a poster child for the tidal disruption of satellite galaxies and the hierarchical build up of galactic stellar haloes. With early observations largely restricted to pencil-beam fields of view \citep{ ivezic00,vivas01,newberg02,martinez-delgado02}, the striking extent of the tidal tails was not revealed until all-sky 2MASS survey became available \citep{majewski03}. It was soon realized that dynamical modelling of the tidal tails provides powerful constraints on the gravitational field of the Milky Way (MW). Yet, after more than two decades of continuous research the Sgr stream still presents a number of theoretical challenges. For example, dynamical models of the Sgr stream that adopt a static MW potential have returned dark matter halo with prolate \citep{helmi04}, oblate \citep{johnston05} and triaxial \citep{law10} shapes. This mismatch may be due to the gravitational attraction induced by a massive Large Magellanic Clound (LMC) on first infall, which has two important effects: it perturbs the orbits of stream stars \citep{law10, vera-ciro13}, and displaces the MW disc from the Galactic barycentre \citep{gomez15,petersen20a}, causing a reflex motion in the kinematics of distant halo stars \citep[][hereafter PP21]{erkal20,pp21} and Local Group galaxies \citep{p16}. Recently, \citet[][hereafter VBE21]{vasiliev20} show that the disparate constraints on the dark matter halo shapes may be due to the adoption of a static MW potential in earlier models. 
However, other issues, like the origin of the bifurcated tails \citep{belokurov06, koposov12}, remain poorly understood. E.g. it was suggested that the bifurcations could be caused by rotation in the progenitor dwarf \citep{p10}. Yet, these models predict a significant amount of rotation in the remnant core, which has been ruled out by different kinematic surveys \citep{p11, frinchaboy12, vasiliev20b, delpino21}. Furthermore, {\it all} published models predict the existence of old stream wraps populated by stars lost during early pericentric passages whose detection remains elusive, probably because they are more phase-mixed than those stripped recently. Finding this material may constrain the amount of dynamical friction experienced by the Sgr dwarf \citep{fardal19b}, and its infall trajectory  \citep{Dierickxloeb17}.

This paper explores a statistical method to detect kinematic substructures in the halo that are partially, or even fully phase mixed. In particular, we use  Gaussian mixture techniques to quantify clustering in angular momentum space. 
Previous attempts to detect the Sgr stream using catalogues of halo stars with full phase-space coordinates either apply cuts in energy and/or angular momentum \citep{li19,johnson20}, or search for stars on similar orbits with Friends-of-Friends algorithms \citep{yang19}. In contrast, {\it our method draws statistical associations between objects in the halo without relying on assumptions on the form, shape or time-evolution of the MW potential, or introducing subjective cuts in integrals of motion}. Its applicability is chiefly limited by the reduced number of halo stars with measured 6D phase-space coordinates. For example, \citet{ramos20} provides a catalogue of 182495 RR Lyrae in {\it Gaia} DR2 \citep{gaia18} with 5D information, that is approximately 17 times larger than the 6D data set introduced below. 
\begin{figure*}
\begin{center}
\includegraphics[width=174mm]{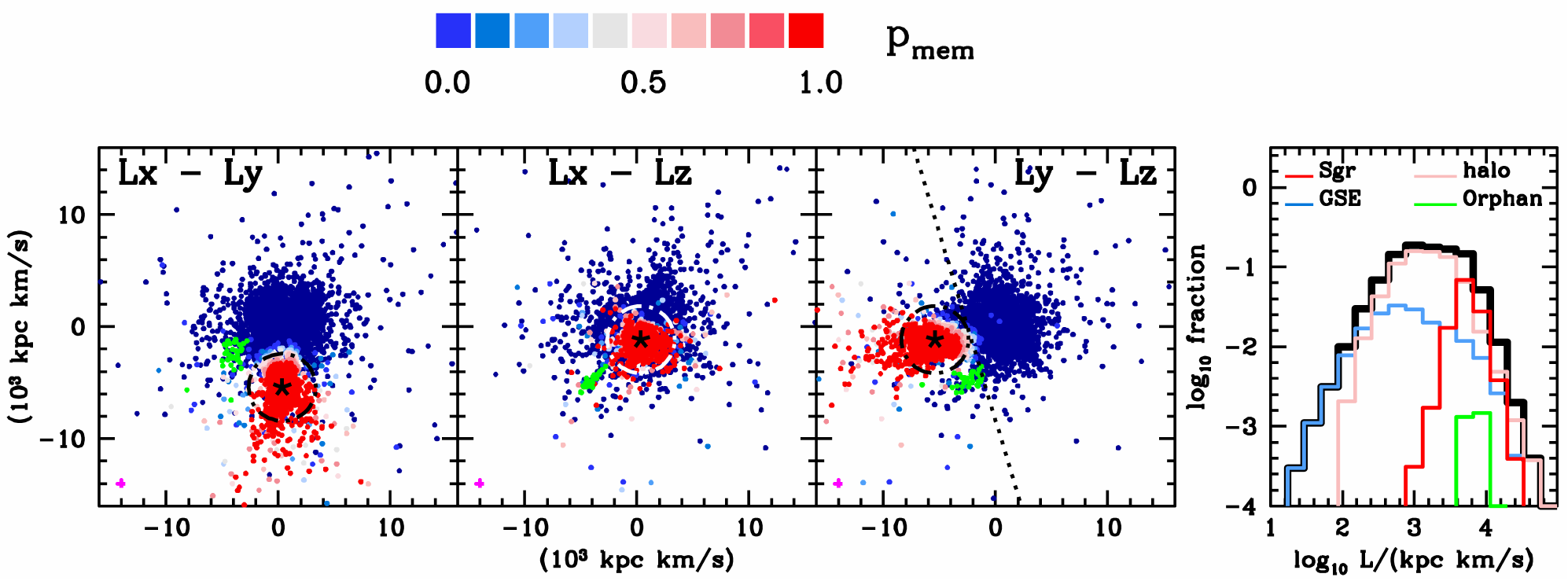}
\end{center}
\caption{Angular momentum components of stellar halo stars with available 6D phase-space coordinates and located within $|B_{\rm sgr}|<20^\circ$ of the Sgr dSph orbital plane (see text). Stars are colour-coded according to their probability to belonging to the Sgr stream. Back stars mark the angular momentum components of the Sgr dwarf galaxy. Black/white circles show the exclusion limits applied by \citetalias{pp21} to remove Sgr stream stars from their analysis. Stars on the left of the dotted line were labelled Sgr stream members by \protect{\citet{johnson20}}. Orphan stream members ($p_{\rm mem}>0.9$) are shown in green. Magenta crosses in the bottom-left of each panel show the average uncertainty of angular momentum measurements, $\sim 600\kpc\kms$. Right panel shows the angular momentum distribution of the stellar sample (black line). Coloured histograms weigh stars by membership probabilities (see text). }
\label{fig:am}
\end{figure*}

\section{Data}\label{sec:data}
We use a parent catalogue of 6030 K Giants and 3952 Blue Horizontal Branch (BHB) stars in the Milky Way stellar halo with available phase-space information compiled by \citetalias{pp21} from public data sets \citep{xue08,yanny09,xue11} and updated with {\it Gaia} EDR3 astrometry \citep{gaia_edr3}. 


One of the main goal of this paper is to identify Sgr stream members. For a better characterization of the angular momentum distribution of the Sgr stream, it is convenient to minimize the fraction of MW interlopers in the Bayesian fits of \S\ref{sec:bayes}. To this aim, the reference frame is rotated such that the average polar angle measured from the Galactic centre is at $B_{\rm sgr}=0$ \citep{majewski03}. Subsequently, stars within $|B_{\rm sgr}|<20^\circ$ at Galactocentric distances $r>20\kpc$ are chosen, which yields 3419 K Giants and 1970 BHBs, with mean distance errors $\epsilon_D/D\simeq 0.05$ and 0.17, respectively. 


In addition, we also use published positions, distances, radial velocities, and proper motions of MW satellites  
\citep{mcconnachie20b} and globular clusters \citep{vasiliev21,baumgardt21} 
in order to explore possible associations with the Sagittarius dwarf. 
To convert heliocentric into Galactocentric quantities we place the sun at \(\vec{x}_\odot =(-8.17,0.0,0.02)\) kpc \citep{gravity19,bennett19}, with \(\vec{v}_\odot =(-12.9,245.6,7.78)\) km s\(^{-1}\) \citep{drimmel18}.

\section{Bayesian analysis} \label{sec:bayes}

An efficient way to model the  distribution of stars in angular momentum $\vec{L}=\vec{r}\times \vec{v}$, with $\vec{r}$ and $\vec{v}$ measured in a Galactocentric frame, is to adopt a Gaussian mixture likelihood \citep{kuhnfeigelson17}
\begin{align}\label{eq:prob}
    \mathcal{L}(\vec{\theta}|\vec{S}) =\sum_{i=1}^{N}\,\alpha_i\,\mathcal {N}_i(\vec{\theta}|\vec{S}),
\end{align}
 with individual weights normalized such that $\sum_{i=1}^{N}\alpha_i=1$. Here, $\vec{\theta}=\{D_i,\ell_i,b_i,v_{\rm los,i},\mu_{\ell,i},\mu_{b,i}\,\epsilon_{D,i},\epsilon_{v,i},\epsilon_{\ell,i},\epsilon_{b,i}\}^{N_{\rm sample}}_{i=1}$  is a sample stars with full phase-space information, and 
 $\vec{S}$ is the array of model parameters that define the multivariate Gaussian probability functions 
\begin{align}\label{eq:gauss}
 \mathcal{N}(\vec{\theta}|\vec{S})=  \frac{1}{[(2\pi)^3|\det(C)|]^{1/2}}\exp\big[-(\vec{L}-\langle \vec{L}\rangle)^T C^{-1}(\vec{L}-\langle \vec{L}\rangle)\big],
\end{align}
where $C$ is the covariance matrix
\begin{align}\label{eq:cov}
C=
\begin{bmatrix}
\epsilon^2_{x}+\sigma_x^2 & \epsilon_{x}\epsilon_y\rho_{xy} & \epsilon_{x}\epsilon_z\rho_{xz} \\
\epsilon_{y}\epsilon_x\rho_{xy} & \epsilon^2_{y} +\sigma_y^2& \epsilon_{y}\epsilon_z\rho_{yz} \\
\epsilon_{z}\epsilon_x\rho_{xz} & \epsilon_{z}\epsilon_y\rho_{yz}& \epsilon^2_{z}+\sigma_z^2
\end{bmatrix} 
\end{align}
and $\rho_{ij}$ are correlation coefficients, while $(\epsilon_x,\epsilon_y,\epsilon_z)$ are uncertainties associated with individual angular momentum values estimated by Monte-Carlo sampling observational errors on heliocentric distance ($\epsilon_D$), line-of-sight velocities ($\epsilon_v$), and proper motions ($\epsilon_\ell,\epsilon_b$). The hyperparameters $\sigma^2_{i}$, with $i=x,y,z$, account for the spread in angular momentum of stars in each Gaussian mixture component beyond those introduced by statistical errors \citep[e.g. see][]{hobson02}. 
For a better exploration of the prior volume, it is convenient to express the mean angular momentum in spherical coordinates $\langle \vec{L}\rangle=(L_x,L_y,L_z)=L(\cos\theta\cos\phi,\cos\theta\sin\phi,\sin\theta)$.
Each component is determined by 7 parameters $\vec{S_i}=\{\log L,\theta,\phi, \sigma_x,\sigma_y,\sigma_z,\alpha\}_i$, with $i=1,...,N$, hence the total number of parameters in our fits is $7 N- 1$. Our analysis uses $N=4$ components with 27 parameters. 

We adopt flat priors on $\{\alpha,\log L,\cos\theta,\phi\}$, and Jeffreys priors for the hyperparameters $\{\sigma_{x}, \sigma_y$, $\sigma_z\}$, with ranges that include reasonable values. 
Mixture models are fitted with the code {\sc MultiNest} \citep{feroz08,feroz09}, which uses a nested-sampling technique \citep{skilling04} to calculate posterior distributions and the evidence of the model.

From the posterior distributions, we estimate the probability that an object (star, GC or dwarf galaxy) with observed quantities $\vec{\theta}$ belongs to the Sgr stream as
\begin{align}\label{eq:pmem}
    p_{\rm mem}(\vec{\theta}) = \frac{\alpha_{\rm sgr}\,\mathcal{N}_{\rm sgr}(\vec{\theta}|\vec{S}_p) }{ \sum_{i=1}^{N}\,\alpha_i \,\mathcal {N}_i(\vec{\theta}|\vec{S}_p)},
\end{align}
where $\vec{S}_p$ is an array that contains the median of the posterior distributions.

\setcounter{table}{0}
\begin{table}
\caption{{Posteriors derived from a Gaussian mixture model with $N=4$ components. Only halo stars within $|B_{\rm sgr}|<20^\circ$ of the Sgr orbital plane and at Galactocentric distances $r>20\kpc$ are included in the fit. Angular momentum $L$ and the hyperparameters $\sigma_i$ are given in units of $\kpc\kms$. } \label{tab:param}}
\begin{center}
\begin{tabular}{|l|c|c|c| c|}
\hline
Param. &  \#1 ~(Sgr) &  \#2 (GSE)  &  \#3 (smooth) &  \#4 (Orphan) \\
\hline
$\alpha$             & $0.322^{+0.002}_{-0.002}$ & $0.239^{+0.002}_{-0.002}$  & $ 0.438^{+0.004}_{-0.005}$ & $0.0077^{+0.0003}_{-0.0003}$ \\
$\langle L_x\rangle$  & $+426^{+7}_{-6}$         & $-0.7^{+0.4}_{-0.4}$       & $+144^{18}_{-19}$    & $-3905^{+20}_{-17}$ \\
$\langle L_y\rangle$  & $-4950^{+11}_{-11}$      & $+11^{+0.3}_{-0.4}$        & $-278^{+10}_{-9}$    & $-2322^{+11}_{-11}$ \\
$\langle L_z\rangle$  & $-1436^{+7}_{-6}$        & $+3^{+0.2}_{-0.2}$         & $+122^{16}_{-16}$    & $-4664^{+19}_{-15}$ \\
$\sigma_x$            & $655^{+5}_{-6}$            & $10.3^{+0.08}_{-0.08}$   & $ 1755^{+8}_{-8}$       & $13^{+1}_{-4}$ \\
$\sigma_y$            & $1255^{+13}_{-14}$         & $11.3^{+0.3}_{-0.4}$     & $ 1926^{+10}_{-12}$     & $12^{+0.6}_{-0.9}$ \\
$\sigma_z$            & $659^{+6}_{-5}$            & $10.2^{+0.06}_{-0.06}$ & $ 1733^{+16}_{-16}$       & $13^{+0.9}_{-1.2}$ \\
\hline
\end{tabular}
\end{center}
\end{table}

\section{Results}\label{sec:results}
\subsection{Halo stars in angular momentum space}
Table~\ref{tab:param} provides the median and 1-sigma uncertainties of the parameters of a Gaussian mixture model with $N=4$ components. 
The posterior distributions on individual parameters are well behaved, with little covariance between them. 
Comparison against the angular momentum of the Sagittarius dwarf, $\vec{L}_{\rm sgr}= \vec{r}_{\rm sgr}\times \vec{v}_{\rm sgr}=(17.9, 2.6, -6.6)\kpc\times (239.5, -29.6, 213.5)\kms = (+360, -5402, -1152)\kpc\kms$ 
\citepalias{vasiliev20}, indicates that the first Gaussian component, with $\alpha\simeq 32\%$ of our sample, corresponds to tidal debris from the Sgr dwarf galaxy. Notice, however, that the median angular momentum of the stream is slightly different from that of the remnant core. This is to be expected given that tidal stripped stars escape from the Lagrange points L1/L2 with lower/higher angular momentum than the disrupting progenitor \citep[e.g.][]{p06}. The hyperparameters, $\sigma_i$, also contain important information. In particular, we find that the spread in the Y-axis, which is roughly perpendicular to the orbital plane of the dwarf, is approximately twice as large as in the other two directions. We come back to this result in \S\ref{sec:dis}.

The remainder components of the Gaussian mixture model also show remarkable properties. For example, Component \#2, with $\sim 24\%$ of the fitted sample, corresponds to stars with little or no angular momentum, thus moving on nearly radial orbits. These properties are similar to those of the Gaia-Sausage-Enceladus (GSE) substructure \citep{belokurov18,helmi18}. 
The properties of this system are worth exploring in a separate contribution. Component \#3 contains the largest fraction of stars with $\simeq 44\%$ of the sample. It also shows the broadest angular momentum distribution, suggesting an association with the extended `smooth' MW halo. Component \#4 is the faintest object in the mixture model with only $\alpha\simeq 0.8\%$ of the sample (roughly 40 stars). The mean angular momentum and the sky coordinates of these stars indicate membership to the Orphan stream \citep{belokurov06,fardal19a}. The detection of such a faint substructure using Gaussian mixture models opens up interesting follow-up applications (see also \S\ref{sec:dis}). We note that adding an extra component ($N=5$) does not lead to the identification of additional substructures.

Fig.~\ref{fig:am} shows projections of the angular momentum vector of 5389 stars in the Sagittarius stream plane ($|B_{\rm sgr}|<20^\circ$) colour-coded according to their probability of belonging to the Sagittarius stream, Equation~(\ref{eq:pmem}). 
For reference, purple crosses on the bottom-left corner mark the average size of individual error bars, $\simeq 600\kpc\kms$. 
This Figure reveals a number of interesting features.
First, we find 171 BHBs and 754 K Giants with a high probability, $p_{\rm mem}>0.9$, of being members of the Sgr stream. As expected, these stars clump around the orbital angular momentum of the Sgr dwarf, which is marked with black stars for reference. However, a considerable amount of scatter is visible in the three panels, specially in the $L_y$ component, which is roughly perpendicular to the orbital plane of the Sgr dwarf. This is consistent with the hyperparameter $\sigma_y$ being approximately twice as large as $\sigma_x$ and $\sigma_z$ (see Table~\ref{tab:param}). Second, the distribution of non-members ($p_{\rm mem}<0.5$) is clearly non-Gaussian. As a result, our mixture models need at least $N=4$ components in order to find a good match. 
The angular momentum $L=|\vec{L}|$ distribution of the sample is shown in the right panel with a black line. To illustrate the location of the Gaussian mixture components we weigh each star by their membership probability and re-compute the histrograms. As expected, we find that stars with high probability of belonging to the GSE substructure have little angular momentum. Stars labelled as members of the `smooth' halo have a broad angular momentum distribution, whereas Sgr and Orphan stars clump in a relatively narrow region.

The angular momentum of the Sgr stream has been recently studied by \citet{johnson20}, who propose a simple criterion for membership. Namely, stars with angular momentum coordinates $L_z< -(2.5+L_y)/0.3$, with $L$ measured in units of $10^{3}\kpc\kms$ (dotted line in Fig.~\ref{fig:am}), are labelled members. Application of this cut to our dataset yields 1500 stars. All the 925 stream members detected in our analysis at $p_{\rm mem}>0.9$ satisfy this condition. We also find 317 likely interlopers ($p_{\rm mem}<0.5$). Hence, these results suggest that \citet{johnson20} sample is complete but has a low purity. Given the results from our sample, where $317/1500\approx 21\%$ of the stars selected by the \citet{johnson20} criterion are low-likelihood members, a similar fraction of the \citet{johnson20} stream sample may be MW interlopers. However, we caution that the \citet{johnson20} catalogue has different sky coverage and phase-space densities, which complicates a direct comparison.

\citetalias{pp21} studied the angular momentum distribution of stars in the outskirts of the MW with the opposite goal in mind, namely to {\it remove} Sgr stream members from a catalogue of `smooth' halo stars. To this aim, they exclude 1062 stars with angular momenta similar to that of the Sgr dwarf, $|\vec{L}-\vec{L}_{\rm sgr}|<3000 \kpc\kms$ (dotted-line black/white circles in Fig.~\ref{fig:am}). One can see by eye that this cut is imperfect, as some Sgr members are located outside the circled regions. Indeed, we find that 793 Sgr stream members ($p_{\rm mem}>0.9$) and 72 MW interlopers ($p_{\rm mem}<0.5$) are located within this volume. This implies that \citetalias{pp21} cut successfully removes a large fraction ($793/925\simeq 86\%$) of stream members from their halo sample, while the misidentification of smooth halo stars as stream members is relatively low, $72/1062\simeq 7\%$.

\begin{figure}
\begin{center}
\includegraphics[width=85mm]{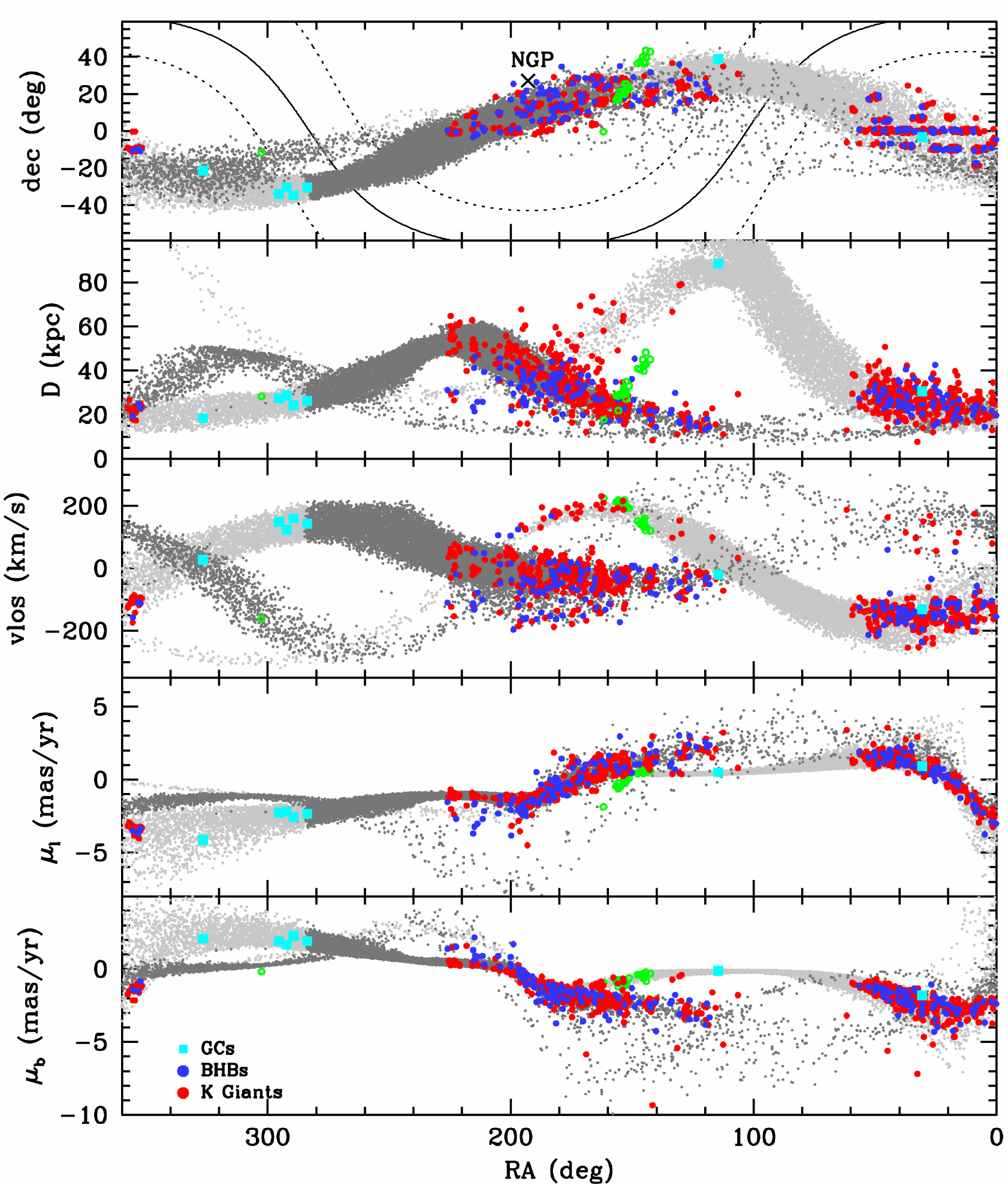}
\end{center}
\caption{Distribution and kinematics of Sgr stream members at a $p_{\rm mem}>90\%$ confidence level (see text). Red and blue dots denote K Giant and BHB stars. Dark/light grey dots show leading/trailing tail particles from \citetalias{vasiliev20} $N$-body model. For reference, the disc plane is shown with a solid black line, with dotted lines showing Galactic latitudes $b=\pm 20^\circ$. The North Galactic Pole is marked with a black cross. Notice the presence of (previously unknown) second wraps of the trailing/leading tails in the Northern (R.A.$>100^\circ$)/Southern (R.A.$<100^\circ$) hemispheres, respectively. We detect 7 Globular Clusters associated with the Sgr stream (cyan squares). Green dots show members of the Orphan stream (component \#4 in Table~\ref{tab:param}).}
\label{fig:dvpm}
\end{figure}

\subsection{The Sagittarius \& Orphan streams}
In Fig.~\ref{fig:dvpm} we plot the observational coordinates of Sgr stream stars with membership probabilities $p_{\rm mem}>0.9$. Blue and red stars denote BHB and K Giant stars, respectively. This plot reveals that while the Sgr stream can be described by a simple multivariate normal function in angular momentum (see Fig.~\ref{fig:am}), the distribution in phase space is extraordinarily more complex. 
To guide the interpretation of the detections, we over-plot with dark/light grey particles the leading/trailing tails of the Sgr stream model recently published by \citetalias{vasiliev20}, which accounts for both the gravitational attraction of the LMC as well as the displacement of the MW disc from the Galactic barycentre in response to the LMC infall \citepalias[see detailed discussion in][]{pp21}. 

 Upper panel of Fig.~\ref{fig:dvpm} shows the sky projections of the stream members. The sky coverage is incomplete, with large gaps at $R.A.\gtrsim 200^\circ$ and $60^\circ\lesssim R.A.\lesssim 100^\circ$ owing to the limited SEGUE footprint. 
 Notice also that all BHBs are located within $r\lesssim 40\kpc$ from the MW centre. This is an observational bias rather than a physical effect, as BHBs are fainter than K Giants, and thus more difficult to detect at large distances. They also have on average larger associated errors in proper motions and radial velocities, thus complicating a statistical association with the stream. 

Comparison against the $N$-body model shows that most stars have distances, line-of-sight velocities and proper motions consistent with those of the most recent wraps of the leading ($120^\circ\lesssim R.A.\lesssim 220^\circ$) and trailing ($-10^\circ\lesssim R.A.\lesssim60^\circ$) tails, respectively. 
In addition, we detect the continuation of the trailing tail in the Northern Hemisphere ($R.A.\sim 170^\circ$ and $v_{\rm los}\sim +200\kms$) as well as the leading tail in the Southern Hemisphere ($R.A.\sim 30^\circ$ and $v_{\rm los}\sim +100\kms$), thus increasing the stream coverage by a factor $\sim 2$ with respect to current 5D maps \citep[e.g.][]{ramos20}. Given that the $N$-body model has not been fitted to this particular data set the overall agreement with observations is remarkable. Yet, a few discrepancies are noticeable. E.g. the continuation of the leading tail at $R.A.\lesssim 60^\circ$ shows systematically lower $v_{\rm los}$ values than predicted by the model. We will return to this issue in \S\ref{sec:dis}. 

Interestingly, material from the old wraps may have been detected by \citet{johnson20} after applying a cut in $L_y$--$L_z$ (see dotted line in Fig.~\ref{fig:am}). These authors found metal-poor stars associated with the Sgr stream that are both off-set and with a diffuse distribution of line-of-sight velocities compared to the metal-rich component, which led them to speculate the possible existence of a stripped stellar halo of the Sagittarius dwarf. However, \S\ref{sec:bayes} shows that the proposed cut in $L_y$--$L_z$ returns a non-negligible number of MW contaminants ($\approx 21\%$ in our sample), suggesting that the broad velocity distribution may be partially explained the presence of MW interlopers. In contrast, selecting members from the Gaussian mixture of Fig.~\ref{fig:dvpm} shows that the older wraps of the trailing/leading tails are kinematically cold, in agreement with \citetalias{vasiliev20} models.

The mixture decomposition presented in \S\ref{sec:bayes} can be used to draw statistical associations between each of the Gaussian components and {\it any} object in our Galaxy with available phase-space information. For example, of the 170 known Globular Clusters in the Milky Way, we find 7 with high probability ($p_{\rm mem}>0.9$) of being members of the Sgr system: M54, Arp 2, Pal 12, Whiting 1, NGC 2419, Terzan 7 and 8 (cyan boxes in Fig.~\ref{fig:dvpm}). The rest have membership probabilities $p_{\rm mem}\lesssim 0.01$, suggesting that there are no additional known GCs associated with the Sgr dwarf, in agreement with  recent analyses \citep[e.g.][]{Arakelyan21,johnson20}. Our results do not support recent claims that NGC 5634 and 4147 trace ancient wraps of the stream \citep{bellazzini20}. No known satellite galaxy appears to be associated with the Sgr dwarf.

In addition, we also detect several members of the Orphan stream (component \#4 in Table~\ref{tab:param}), shown in Fig.~\ref{fig:dvpm} with green dots. Comparison against the positions, velocities and proper motions found by \cite{fardal19b} shows excellent agreement, indicating that the Gaussian mixture decomposition of \S\ref{sec:bayes} can also uncover faint substructures in the stellar halo even when these contribute to $\lesssim 1\%$ of the fitted sample.

\section{Discussion \& Summary}\label{sec:dis}
We show that modelling the distribution of stars in angular momentum space with Gaussian mixture techniques provides a powerful method to detect accreted substructures that are partially, or fully mixed in phase-space, without making assumptions on the form, shape or time-evolution of the MW potential. Application to a catalogue of 5389 stars in the plane of the Sgr stream with available 6D phase-space coordinates reveals the presence of at least four independent dynamical components. We associate the broadest and most numerous with the `smooth' stellar halo, and the faintest (with only $0.8\%$ of the sample) with the Orphan stream. Interestingly, we also identify a third component with little or no angular momentum that likely corresponds to the GSE substructure.

\begin{figure}
\begin{center}
\includegraphics[width=82mm]{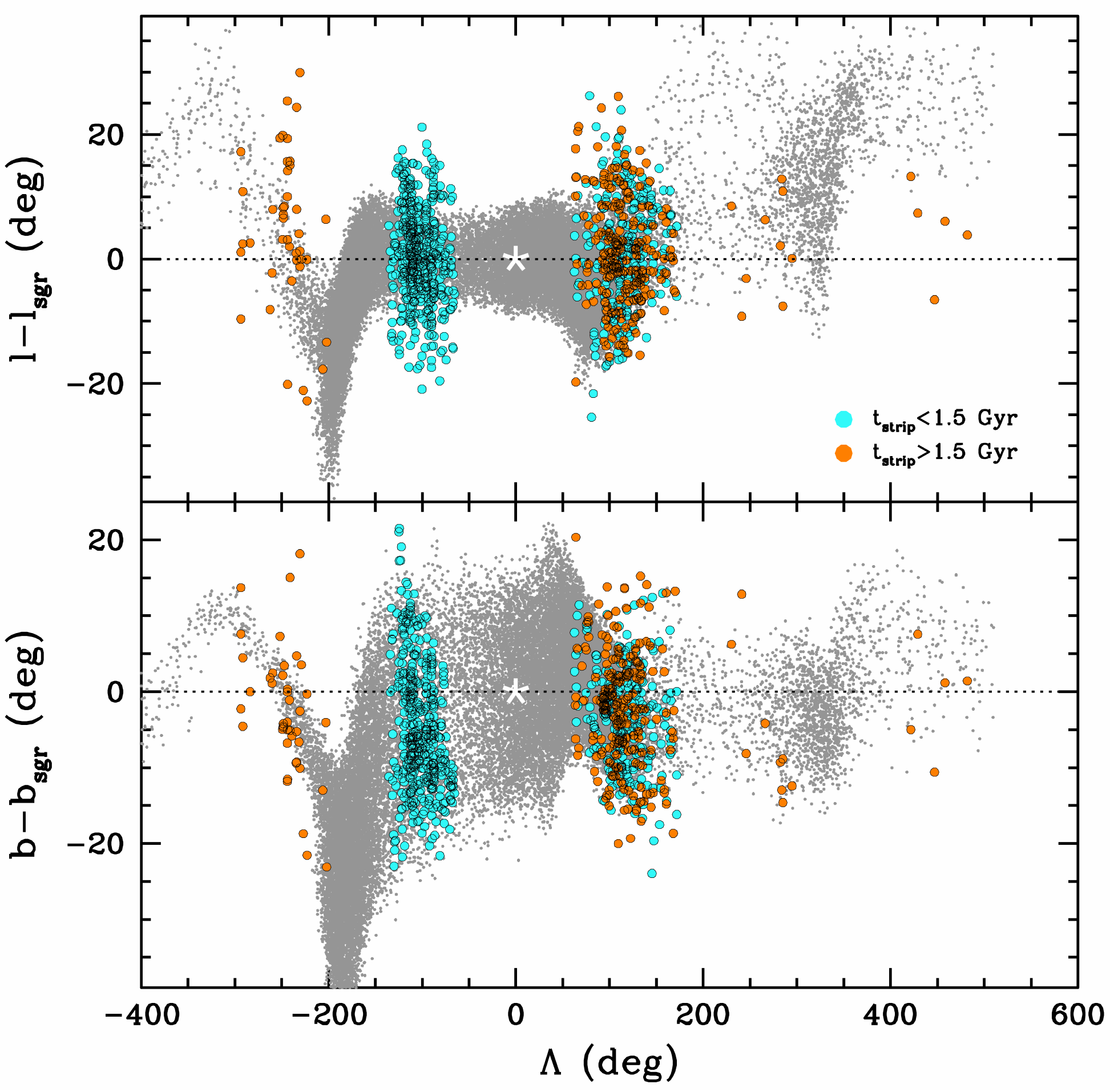}
\end{center}
\caption{Galactocentric coordinates of the orbital poles of individual stream members ($p_{\rm mem}>0.9)$ relative to that of the Sgr dwarf as a function of the angle along the stream, $\Lambda$. Stars are colour-coded according to the time when they were tidally stripped from the progenitor (see text). The location of the Sgr dwarf is marked with a white star.}
\label{fig:L_plane}
\end{figure}

Our statistical technique detects two older wraps of the Sagittarius stream that correspond to the continuation of the leading/trailing tails in the Southern/Northern hemispheres, 
showing that the Sgr stream circles the Galaxy at least {\it twice}. The full extent of the tidal tails is shown in Fig.~\ref{fig:L_plane} in a reference frame aligned with the orbital plane of the stream. 
Dots show the orbital poles of individual stream members relative to that of the Sagittarius dwarf galaxy as a function of the angular separation from the remnant core (with notation $\Lambda>0$ for the leading, and $\Lambda<0$ trailing tails). The discovery of older wraps double the known extent of the Sgr stream. While published 5D data cover parts of the stream within $-150^\circ\lesssim \Lambda\lesssim 200^\circ$ from the Sgr dwarf, 6D detections increase this range out to $-300^\circ\lesssim \Lambda\lesssim 500^\circ$.

For comparison, the orbital poles of the $N$-body model are shown with grey dots. To estimate the time at which different parts of the stream became tidally unbound from the Sgr dwarf, we find the $N$-body particle closest to each individual star and colour-code stream members according to their stripping time. We distinguish between material lost at $t_{\rm strip}<1.5\Gyr$ (cyan) and $t_{\rm strip}>1.5\Gyr$ (orange), which approximately correspond to particles unbound during the last and penultimate pericentric passages of the progenitor dwarf, respectively \citepalias[for details, see][]{vasiliev20}. As expected, the continuation of the leading \& trailing tails correspond to material stripped early. Interestingly, the orbital poles of the old wraps remain roughly aligned with those of the Sgr dwarf. This appears somewhat in tension with the $N$-body model, which predicts a strong bending of the leading tail at $\Lambda\gtrsim +400^\circ$. The mismatch is mainly due to the systematically higher line-of-sight velocities predicted for the continuation of the leading tail at $R.A.\lesssim 60^\circ$ (see Fig.\ref{fig:dvpm}).
Notice also that the orbital poles are considerably more scattered in the longitudinal direction (which is roughly perpendicular to the orbital plane) than predicted by the model. This may be connected with the notoriously large spread of the component $L_y$ highlighted in Fig.~\ref{fig:am} and quantified in Table~\ref{tab:param} via the hyperparameter $\sigma_y$. The large scatter of the orbital poles and $L_y$ values may be related to the bifurcation of the leading \& trailing tails \citep[][]{yang19}. 
The  detection of old stream wraps provide new constraints to investiate the origin of the bifurcation in more detail.

The Gaussian mixture decomposition identifies $\sim 40$ members of the Orphan stream, which contribute to $\approx 0.8\%$ of the fitted sample. It thus appears that our statistical analysis is able to detect narrow 
clumps in angular momentum with a small number of members --which are typically associated with disrupting globular clusters or faint satellite galaxies-- as well as fully phase-mixed, hot substructures lurking in the stellar halo. 
Follow-up contributions will explore whether adding metallicity and/or chemical elements as independent dimensions in the Bayesian likelihood~(eq. \ref{eq:prob}) helps to uncover clustering in angular momentum at even finer detail. 

\section*{Acknowledgements}
We thank Teresa Antoja, Pau Ramos, Eugene Vasiliev, Vasily Belokurov, Denis Erkal, Benjamin Johnson and Charlie Conroy for helpful comments. MSP acknowledges funding from a UK Science and Technology Facilities Council (STFC) Consolidated Grant.

\section*{Data Availability}
The Sgr stream sample may be found on github: \url{https://github.com/michael-petersen/SgrL}. 

\bibliography{reflex}

\end{document}